\documentclass[12pt]{article}
\usepackage{a4,amsthm,amsfonts,latexsym}

\newtheorem{thm}{THEOREM}[]

\newtheorem{lem}{LEMMA}
\theoremstyle{definition}
\newtheorem{rem}{Remark}

\def\mfr#1/#2{\hbox{${{#1} \over {#2}}$}}
\newcommand{\beq}{\begin{equation}}
\newcommand{\eeq}{\end{equation}}

\newcommand{\bsigma}{\mathord{\hbox{\boldmath $\sigma$}}}
\newcommand{\rmi}{i}

\newcommand{\rmd}{d}

\newcommand{\OO}{O}
\newcommand{\infspec}{{\rm inf\ spec\ }}
\newcommand{\A}{{\bf A}}
\newcommand{\B}{{\bf B}}
\newcommand{\0}{{\bf 0}}

\newcommand{\x}{{\bf x}}

\newcommand{\y}{{\bf y}}

\newcommand{\R}{{\mathbb R}}
\newcommand{\C}{{\mathbb C}}

\newcommand{\half}{\mbox{$\frac{1}{2}$}}

\newcommand{\Ll}{{\mathcal L}}

\newcommand{\eps}{\varepsilon}

\newcommand{\HN}{H_{N,Z,\A}}
\newcommand{\sig}{\bsigma}

\date{\small June 5, 2000}

\begin{document}
\title{\bf{On the maximal ionization of atoms \\ in strong magnetic
fields}}
\author{\vspace{5pt} Robert Seiringer\\
\vspace{-4pt}\small{Institut f\"ur Theoretische Physik,
Universit\"at Wien}\\ \vspace{-4pt} \small{Boltzmanngasse 5,
A-1090 Vienna, Austria}\\ \small{E-Mail:
\texttt{rseiring@ap.univie.ac.at} }}

\maketitle

\begin{abstract}
We give upper bounds for the number of spin $\half$ particles that
can be bound to a nucleus of charge $Z$ in the presence of a
magnetic field $\B$, including the spin-field coupling. We use Lieb's 
strategy, 
which is known to yield $N_c<2Z+1$ for magnetic fields that go to zero at 
infinity, ignoring the spin-field interaction. For particles with fermionic 
statistics in a homogeneous magnetic field our upper bound has an additional 
term of order $Z\times\min\left\{(B/Z^3)^{2/5},1+|\ln(B/Z^3)|^2\right\}$.
\end{abstract}

\section{Introduction and main result}

Let $\HN$ be the Hamiltonian for $N$ identical particles with spin
$\half$ in the Coulomb field of a nucleus of charge $Z$ and in a
magnetic field $\B=\rm{curl\,} \A$,
\beq\label{ham}
\HN=\sum_{i=1}^N\left(H_\A^{(i)}-\frac
Z{|\x_i|}\right)+\sum_{i<j}\frac1{|\x_i-\x_j|},
\eeq
with
\beq
H_\A^{(j)}=\left[\bsigma_j\cdot\left(-\rmi\nabla_j+
\A(\x_j)\right)\right]^2=
\left(-\rmi\nabla_j+\A(\x_j)\right)^2+\B\cdot\sig_j.
\eeq
Here $\A\in\Ll^2_{\rm loc}(\R^3;\R^3)$ is the magnetic potential
and $\sig=(\sigma_x,\sigma_y,\sigma_z)$ are the usual Pauli
matrices. The Hamiltonian (\ref{ham}) acts on the fermionic
(respectively bosonic) subspace of
$\bigotimes^N\Ll^2(\R^3,\rmd\x;\C^2)$. We will assume that the
ground state energy 
\beq
E(N,Z,\B)=\infspec\HN
\eeq
is finite. (Note that the energy depends only on $\B$ because of gauge 
invariance.) A sufficient condition for this is $\B\in\Ll^{3/2}+\Ll^\infty$, 
because 
is this case $|\B|+Z/|\x|$ is relatively bounded with respect to $-\Delta$, 
and
$\infspec\HN\geq \infspec\sum_{i=1}^N\left( 
-\Delta_i-|\B(\x_i)|-Z/|\x_i|\right)$ by the 
diamagnetic inequality.
Moreover,
we will only consider magnetic potentials $\A$ such that the energy
$E(N,Z,\B)$ is {\it monotonously decreasing} in $N$ for fixed $Z$.
This is in particular the case for a homogeneous magnetic field. We
are interested in the maximal number of particles that can be
bound, i.e. the largest $N$ such that $E(N,Z,\B)$ is an
eigenvalue. We will denote this \lq\lq critical\rq\rq\ $N$ by
$N_c$, suppressing the dependence on $Z$ and $\B$. For simplicity
we will restrict ourselves to considering identical particles.

Alternatively, one could define the critical particle number by
\beq
\widetilde N_c=\max\left\{N|E(N,Z,\B)<E(N-1,Z,\B)\right\}.
\eeq
With this definition $E(N,Z,\B)$ is certainly an eigenvalue if $N\leq 
\widetilde N_c$, so $\widetilde N_c\leq N_c$. Hence any upper bound to $N_c$ 
is also an upper bound to $\widetilde N_c$. 

It is well known that magnetic fields, at least
homogeneous ones, enhance binding. In
\cite{ahs3} it is shown that every once negatively charged ion (i.e.
$N=Z+1$) has an infinite number of bound states in the presence of a
homogeneous magnetic field of arbitrary field strength $B$. And in
\cite{lsya} the lower bound $\liminf_{Z\to\infty}(\widetilde N_c/Z)\geq 2$ as
long as $B/Z^3\to\infty$ is given, which is in contrast to asymptotic
neutrality in the absence of magnetic fields \cite{lsst}.

We will use Lieb's strategy \cite{lieb} to derive an upper bound
on $N_c$. The difference between our considerations and
\cite{lieb} is the coupling of the spin to the magnetic field,
i.e. the term $\B\cdot\sig$ in the Hamiltonian. Without this term,
Lieb derived the bound $N_c<2 Z+1$ for any bounded $\A$ that goes to zero 
at infinity. 

Our result is as follows:
\begin{thm}[Upper bound on $N_c$]\label{thm1}
Under the conditions stated above,
\beq\label{arrive}
N_c< 2Z+1+\frac12\frac{E(N_c,Z,\B)-E(N_c,kZ,\B)}{N_cZ(k-1)}
\eeq
for all values of $k>1$.
\end{thm}

Note that since the ground state energy is superadditive in $N$,
$E(N,Z,\B)/N$ is bounded by some some function independent of $N$.
Moreover, the best bound in (\ref{arrive}) is achieved in the
limit $k\searrow 1$, which exists by concavity of $E(N,Z,\B)$ in
$Z$.

To apply Theorem \ref{thm1} to the case of fermionic electrons in a 
homogeneous magnetic field $\B=(0,0,B)$, one needs upper and lower bounds to 
the ground state energy. These were derived in \cite{lsya} and \cite{lsyb} and 
are given in 
section \ref{sf}. The result is the following:

\begin{thm}[Maximal ionization for fermions]\label{fermi} Let $\HN$ be the 
restriction of (\ref{ham}) to the fermionic subspace, and let $\B=(0,0,B)$.
Then, for some constants $C_1$ and $C_2$, and for all values of $B\geq 0$
and $Z>0$,
\beq\label{arferm}
N_c< 2Z+1+C_1 Z^{1/3}+C_2 
Z\min\left\{(B/Z^3)^{2/5},1+|\ln(B/Z^3)|^2\right\}.
\eeq
\end{thm}

Of course we do not believe that these bounds are optimal.
One might assume that Lieb's bound $N_c<2Z+1$ holds also in this case, at 
least for large $Z$ (compare with the lower bound in \cite{lsya} stated 
above), but it remains an open problem to show this. However, Theorem 
\ref{fermi} improves a result obtained in \cite{brummel}, which states that 
$N_c<2Z+1+cB^{1/2}$ for the Hamiltonian (\ref{ham}) restricted to some special 
wave functions in the lowest Landau band, which reduces the problem to an 
essentially one-dimensional one.

One might ask how the Pauli principle affects the result in Theorem 
\ref{fermi}. It turns out that the analogue for bosonic particles is the 
following:

\begin{thm}[Maximal ionization for bosons]\label{bose}
Let $\HN$ be the restriction of (\ref{ham}) to the bosonic subspace, and let 
$\B=(0,0,B)$.
Then for some constant $C_3$ and for all $B\geq 0$ and $Z>0$ 
\beq\label{26}
N_c< 2Z +1 +\frac Z2\min\left\{\left(1+ \frac 
B{Z^2}\right),C_3\left(1+\left[\ln\left(\frac
B{Z^2}\right)\right]^2 \right)\right\}.
\eeq
\end{thm}

In the next section we will give the proof of Theorem \ref{thm1}. In section 
\ref{sg} several possible generalizations are stated, and in section \ref{se} 
the necessary energy bounds for the case of a homogeneous magnetic field are 
given, which will prove Theorems \ref{fermi} and \ref{bose}.

\section{Proof of Theorem \ref{thm1}}

Let $\Psi$ be a normalized ground state for $\HN$. Assume, for the moment,
that $\langle\Psi||\x_N|\Psi\rangle$ is finite. Then
\begin{eqnarray}\nonumber
& &E(N,Z,\B)\langle|\x_N|\Psi|\Psi\rangle=\langle|\x_N|\Psi|\HN\Psi\rangle\\
\nonumber & &= \langle|\x_N|\Psi| \left(H_{N-1,Z,\A}+H_\A^{(N)}-\frac
Z{|\x_N|}+\sum_{i=1}^{N-1}\frac
1{|\x_i-\x_N|}\right)\Psi\rangle.\\ \label{hn1}
\end{eqnarray}
Because $\Gamma=\int \Psi^*\Psi |\x_N|d\x_N$ is an acceptable trial
density matrix, we can use the variational principle to conclude that
\beq
\langle|\x_N|\Psi|H_{N-1,Z,\A}|\Psi\rangle\geq
E(N-1,Z,\B)\langle|\x_N|\Psi|\Psi\rangle.
\eeq
By assumption the energy is monotonously decreasing in $N$, so
\begin{eqnarray}\nonumber
& &\langle|\x_N|\Psi|\left(H_\A^{(N)}-\frac Z{|\x_N|}+\sum_{i=1}^{N-1}\frac
1{|\x_i-\x_N|}\right)\Psi\rangle\\ \label{eq1} & &\leq
\left(E(N,Z,\B)-E(N-1,Z,\B)\right)\langle|\x_N|\Psi|\Psi\rangle\leq
0.
\end{eqnarray}
Now using the demanded symmetry of $\Psi$, we get
\beq
\langle\Psi| \frac{|\x_N|}{|\x_i-\x_N|}\Psi\rangle=\frac12
\langle\Psi| \frac{|\x_N|+|\x_i|}{|\x_i-\x_N|}\Psi\rangle >
\frac12
\eeq
(the strict inequality follows from the fact that $\left\{(\x,\y),
|\x-\y|=|\x|+|\y|\right\}$ has measure zero), so (\ref{eq1}) gives
\beq\label{eq2}
Z> \frac12 (N-1)+\langle|\x_N|\Psi|H_\A^{(N)}|\Psi\rangle.
\eeq
As in \cite{brummel} we have for any positive function
$\varphi(\x_N)$
\begin{eqnarray}\nonumber
\langle\varphi\Psi|H_\A^{(N)}|\Psi\rangle
&=&\langle\varphi^{1/2}\Psi|\left(H_\A^{(N)}-\left|
\frac{\nabla\varphi}{2\varphi}\right|^2\right)
\varphi^{1/2}\Psi\rangle\\ \label{brum} & &-\rmi{\rm Re\,}
\langle\varphi^{1/2}\Psi|
\frac{\nabla\varphi}{\varphi}\cdot(-\rmi\nabla+\A)
\varphi^{1/2}\Psi\rangle.
\end{eqnarray}
Now $\langle|\x_N|\Psi|H_\A^{(N)}\Psi\rangle$ is certainly real,
because all the other quantities in equation (\ref{hn1}) are real. So
choosing $\varphi(\x_N)=|\x_N|$ in equation (\ref{brum}) we get
\beq\label{real}
\langle|\x_N|\Psi|H_\A^{(N)}\Psi\rangle=
\langle|\x_N|^{1/2}\Psi|\left(H_\A^{(N)}-\frac 1{4|\x_N|^2} \right)
|\x_N|^{1/2}\Psi\rangle.
\eeq
Using that $H_\A^{(N)}\geq 0$ equation (\ref{eq2}) reads
\beq
N< 2 Z+1+\frac 12\langle\Psi||\x_N|^{-1}\Psi\rangle.
\eeq
Moreover, 
\begin{eqnarray}\nonumber
\langle\Psi||\x_N|^{-1}\Psi\rangle(k-1)&=&\frac 1{NZ}\langle
\Psi|(\HN-H_{N,kZ,\A})\Psi\rangle\\ &\leq& \frac
1{NZ}\left(E(N,Z,\B)-E(N,kZ,\B)\right),
\end{eqnarray}
so we arrive at the desired bound for $N_c$.

Throughout, we have assumed that $\langle|\x_N|\Psi|\Psi\rangle$
is finite. A priori, this need not be the case. However, one
could arrive at the same conclusions using the bounded function
$\varphi_\eps(\x_N)=|\x_N|(1+\eps|\x_N|)^{-1}$ instead of $|\x_N|$
in (\ref{hn1}), and letting $\eps\to 0$ at the end (see \cite{lieb}).
Note that
\beq
\left|\frac{\nabla\varphi_\eps}{\varphi_\eps}\right|^2=
\frac1{|\x|^2(1+\eps|\x|)^2}\leq\frac 1{|\x|\varphi_\eps(\x)},
\eeq
so our conclusions remain valid.

\begin{rem}
Instead of ignoring the kinetic energy in (\ref{real}) one could
use the operator inequality
\beq
(-\rmi\nabla+\A)^2-\frac 1{4|\x|^2}\geq 0
\eeq
to conclude that
\beq
N_c< 2Z+1-2\langle\Psi||\x_N|\B(\x_N)\cdot\sig_N\Psi\rangle.
\eeq
This may especially be of interest if $|\B(\x)|\leq b|\x|^{-1}$
for some constant $b$. And for $\B=\0$ Lieb's bound $N_c<2Z+1$ is reproduced.
\end{rem}

\section{Generalizations of Theorem \ref{thm1}}\label{sg}

As in \cite{lieb} several generalizations of Theorem \ref{thm1} are
possible:
\begin{itemize}
\item One can allow different statistics than the bosonic or fermionic
one, or even consider independent particles. Moreover, the particles could
have different masses and charges.
\item Hartree- and Hartree-Fock theories can be treated in the same
manner.
\item One can replace the Coulomb interaction (everywhere) by some
positive $v(\x)=1/w(\x)$, with $w$ satisfying
\beq
w(\x-\y)\leq w(\x)+w(\y),
\eeq
and for some constant $C$
\beq
\left|\nabla w\right|^2\leq C .
\eeq
Looking at the proof of Theorem \ref{thm1} we see that these two
properties are really what we needed.
\end{itemize}

\section{Application to homogeneous fields}\label{se}

We will now apply Theorem \ref{thm1} to the case of a homogeneous
magnetic field $\B=(0,0,B)$, and prove Theorems \ref{fermi} and \ref{bose}.  
The magnetic potential, in the symmetric gauge, is given by $\A=\half 
\B\times\x$. The
energy in this case will be
denoted by $E(N,Z,B)\equiv E(N,Z,\B)$. To derive explicit bounds
on $N_c$, we need upper and lower bounds to the ground state energy of
(\ref{ham}). However, since we are not trying to give the optimal
constants,
the upper bound $E(N,Z,B)\leq 0$ will suffice for our purposes. So we will 
concentrate on the lower bounds. We
will distinguish between the fermionic and the bosonic case.
Throughout, every fixed constant will be denoted by $C$, although
the various constants may be different.

\subsection{The fermionic case}\label{sf}
In \cite{lsya} and \cite{lsyb} the following lower bounds on the
ground state energy of ({\ref{ham}) were derived:
\begin{lem}[Lower bounds on the fermionic energy]
Let $\lambda=N/Z$. The ground state energy of (\ref{ham})
restricted to the fermionic subspace satisfies:
\begin{enumerate}
\item[(a)] If $B\leq C Z^{4/3}$ then
\beq
E(N,Z,B)\geq -CZ^{7/3}\lambda^{1/3}\left(1+C\lambda^{2/3}\right).
\eeq
\item[(b)] If $B\geq C Z^{4/3}$ then
\beq\label{useful}
E(N,Z,B)\geq
-CZ^{9/5}\lambda^{3/5}B^{2/5}\left(1+C\lambda^{-2/5}\right).
\eeq
\item[(c)] If $B\geq C Z^2$ then
\beq\label{only}
E(N,Z,B)\geq -CNZ^2\left(1+\left[\ln\left(\frac C{\lambda^{1/2}}
\left(\frac B{Z^3}\right)^{1/2}+1\right)\right]^2\right).
\eeq
\end{enumerate}
\end{lem}

\begin{rem}
Part (c) of Theorem \ref{fermi} follows from omitting the
repulsion terms in Theorem 1.2 (\lq\lq Confinement to the lowest Landau 
band\rq\rq) in \cite{lsya} and then using the
bound (4.11) there. Although this theorem is applicable for $B\geq CZ^{4/3}$, 
with an additional error term, the result is simpler for $B\geq CZ^2$. 
However, the bound (\ref{only}) is only of interest for $B\geq CZ^3$, because 
for smaller $B$ (\ref{useful}) is more useful.
\end{rem}

Using (\ref{arrive}) and the
bounds in the preceding Lemma we find that $\lambda_c\equiv N_c/Z$ satisfies
\begin{eqnarray}\nonumber
\lambda_c&<&
2+Z^{-1}+C\lambda_c^{-2/3}Z^{-2/3}\left(1+C\lambda_c^{2/3}\right)
\quad {\rm if}\quad B\leq C Z^{4/3},\\ \nonumber \lambda_c&<&
2+Z^{-1}+C\lambda_c^{-2/5}\left(\frac
B{Z^3}\right)^{2/5}\left(1+C\lambda_c^{-2/5}\right)\quad  {\rm
if}\quad B\geq C Z^{4/3}, \\ \nonumber \lambda_c&<&
2+Z^{-1}+C\left(1+\left[\ln\left(\frac C{\lambda_c^{1/2}}
\left(\frac B{Z^3}\right)^{1/2}+1\right)\right]^2\right)\quad {\rm
if}\quad B\geq C Z^{2}.\\
\end{eqnarray}
Putting together these three bounds, we obtain the result stated in Theorem 
\ref{fermi}. Note that these bounds imply in particular
\beq
\limsup_{Z\to\infty}\frac {N_c}Z\leq 2\qquad {\rm if}\quad B/Z^3\to 0.
\eeq

\subsection{The bosonic case}\label{sb}
To get a lower bound the bosonic energy we will first omit the positive 
repulsion terms in (\ref{ham}). By scaling the variables $\x_i\to Z^{-1}\x_i$ 
we see that
\beq
E(N,Z,B)\geq N Z^2 e(B/Z^2),
\eeq
where $e(b)$ is the ground state energy of hydrogen in a
homogeneous magnetic field of strength $b$. For small $b$, we will use the
diamagnetic inequality, which implies
\beq
e(b)\geq -\frac14-b.
\eeq
A large $b$ expansion of $e(b)$ is given in \cite{ahs3}. From
there we get the following lower bound:

\begin{lem}[Lower bound for the hydrogen energy]
For large enough $b$ the ground state energy of
hydrogen, $e(b)$, satisfies
\beq
e(b)\geq -\frac14\left(\ln b\right)^2\left(1+\frac C{\ln b}\right).
\eeq
\end{lem}

Note that the bosonic energy is at least of order $NZ^2$, even for small $B$. 
Therefore the contribution to (\ref{arrive}) is always at least $\OO(Z)$, in 
contrast to fermions, where the energy is of order $N^{1/3}Z^2$ for small $B$ 
(this is the reason for the additional factor $Z^{1/3}$ in (\ref{arferm})). 
Setting $k=2$ we arrive at the bound given in Theorem \ref{bose}.

We remark that since the bosonic energy is always less than the
fermionic energy, Theorem \ref{bose} holds also for fermions; but the bound 
stated there
is certainly worse than the one given in Theorem \ref{fermi}.

\end{document}